\documentclass[9pt,twocolumn,twoside]{opticajnl}
\journal{opticajournal} 

\setboolean{shortarticle}{true}

\usepackage{lineno}
\usepackage{pifont}

\def\R{\mathbb{R}}

\def\P{\mathbb{P}}
\def\r{\mathbf{r}}
\def\t{\mathbf{t}}
\def\E{\textrm{E}}

\def\F{\mathcal{F}}
\def\N{\mathcal{N}}
\def\bth{\boldsymbol{\theta}}

\def\g{g^*}

\def\e{\mathrm{e}}
\def\j{\mathrm{j}}

\def\k{\mathbf{k}}
\def\n{\mathbf{n}}
\def\0{\mathbf{0}}
\def\d{\mathrm{d}}
\usepackage{mathtools}

\title{Point-Spread-Function Engineering in MINFLUX: Optimality of Donut and Half-Moon Excitation Patterns}

\author[1]{Yan Liu}
\author[1]{Jonathan Dong}
\author[2,3,4]{Juan Augusto Maya}
\author[2]{Francisco Balzarotti}
\author[1, *]{Michael Unser}

\affil[1]{Biomedical Imaging Group, \'Ecole polytechnique f\'ed\'erale de Lausanne, Station 17, 1015 Lausanne, Switzerland}
\affil[2]{Research Institute of Molecular Pathology (IMP), Vienna BioCenter (VBC), Campus-Vienna-Biocenter 1, 1030 Vienna, Austria}
\affil[3]{Facultad de Ingeniería, Universidad de Buenos Aires, Av. Paseo Col\'on 850, C1063ACV Buenos Aires, Argentina}
\affil[4]{Centro de Simulación Computacional (CSC), Consejo Nacional de Investigaciones Científicas y Técnicas (CONICET), Godoy Cruz 2390, C1425FQD Buenos Aires, Argentina} 

\affil[*]{Corresponding author: michael.unser@epfl.ch}

\begin{abstract}
Localization microscopy enables imaging with resolutions that surpass the conventional optical diffraction limit. 
Notably, the MINFLUX method achieves super-resolution by shaping the excitation point-spread function (PSF) to minimize the required photon flux for a given precision.
Various beam shapes have recently been proposed to improve localization efficiency, yet their optimality remains an open question. 
In this work, we deploy a numerical and theoretical framework to determine optimal excitation patterns for MINFLUX.
Such a computational approach allows us to search for new beam patterns in a fast and low-cost fashion, and to avoid time-consuming and expensive experimental explorations.
We show that the conventional donut beam is a robust optimum when the excitation beams are all constrained to the same shape. 
Further, our PSF engineering framework yields two pairs of half-moon beams (orthogonal to each other) which can improve the theoretical localization precision by a factor of about two. 
\end{abstract}

\setboolean{displaycopyright}{false} 

\begin{document}

\maketitle
Fluorescent super-resolution microscopy techniques such as stimulated-emission-depletion microscopy \cite{hell1994breaking}, stochastic optical reconstruction microscopy \cite{rust_sub-diffraction-limit_2006}, and photo-activated localization microscopy \cite{betzig_imaging_2006} 
break the diffraction limit of light by manipulating the emission state of fluorophores \cite{hell_far-field_2007}, achieving resolutions as low as few nanometers. 
By virtue of localizing single molecules, they are widely used in the study of the structure and function of proteins in life sciences \cite{Sillibourne2011, Scarselli2016, Veeraraghavan2016, masch2018, vavrdova_multicolour_2019}. 
 
Maximally INFormative Luminescence eXcitation (MINFLUX) is an emergent localization strategy that reduces the number of photons required to achieve a given localization precision in comparison to camera-based approaches \cite{balzarotti_nanometer_2017}. 
It relies on utilizing tailored excitation beam patterns sequentially to probe the location of a fluorophore.
MINFLUX achieves single-digit nanometer isotropic localization precision in three dimensions of fixed and living cells \cite{gwosch_minflux_2020} and sub-millisecond time resolution for single-molecule tracking \cite{eilers_minflux_2018}.
Using minimal number of photons, MINFLUX reduces photobleaching and enables long observation time, which makes it a suitable technique to study cellular dynamics at the molecular level \cite{Wirth2023, carsten_minflux_2023}. 

Information theory and, in particular, the Cramér-Rao bound (CRB) are commonly used to quantify the precision of localization microscopy \cite{shechtman_optimal_2014, tzang_two-photon_2019, nehme_deepstorm3d_2020, dong2021fundamental, Fu2022, opatovski_depth-enhanced_2024} or in the more general context of optical imaging  \cite{bouchet2021fundamental}. 
This quantity can then be used as a performance metric to optimize and discover new beam shapes in PSF engineering, e.g. for 3D localization in a wide field configuration \cite{shechtman_optimal_2014}.
To the best of our knowledge, there has been no previous work applying such a framework to MINFLUX. 

In this Letter, we introduce an iterative optimization procedure and a theoretical study to determine optimal beam configurations for MINFLUX. 
Our framework is based on a modular pipeline that takes as input the pupil function parameterized by Zernike coefficients, then models light propagation to the sample, and finally outputs the average CRB over a defined region of interest.  
The pupil function is optimized by an iterative gradient-descent-type algorithm, leveraging the automatic differentiation capability of PyTorch, a popular deep-learning library.
We first confirm that the donut beam is an optimal excitation PSF under the constraint of a single excitation beam shape;
we then show that two pairs of orthogonal half-moon beams lead to a further increase in the localization precision by a factor of two if more than one beam shape is possible, which has recently been implemented experimentally \cite{geismann2023fast,deguchi2024simple}. 

\begin{figure*}[tb]
    \centering
    \begin{minipage}[b]{1\linewidth}
      \centerline{\includegraphics[width=\linewidth]{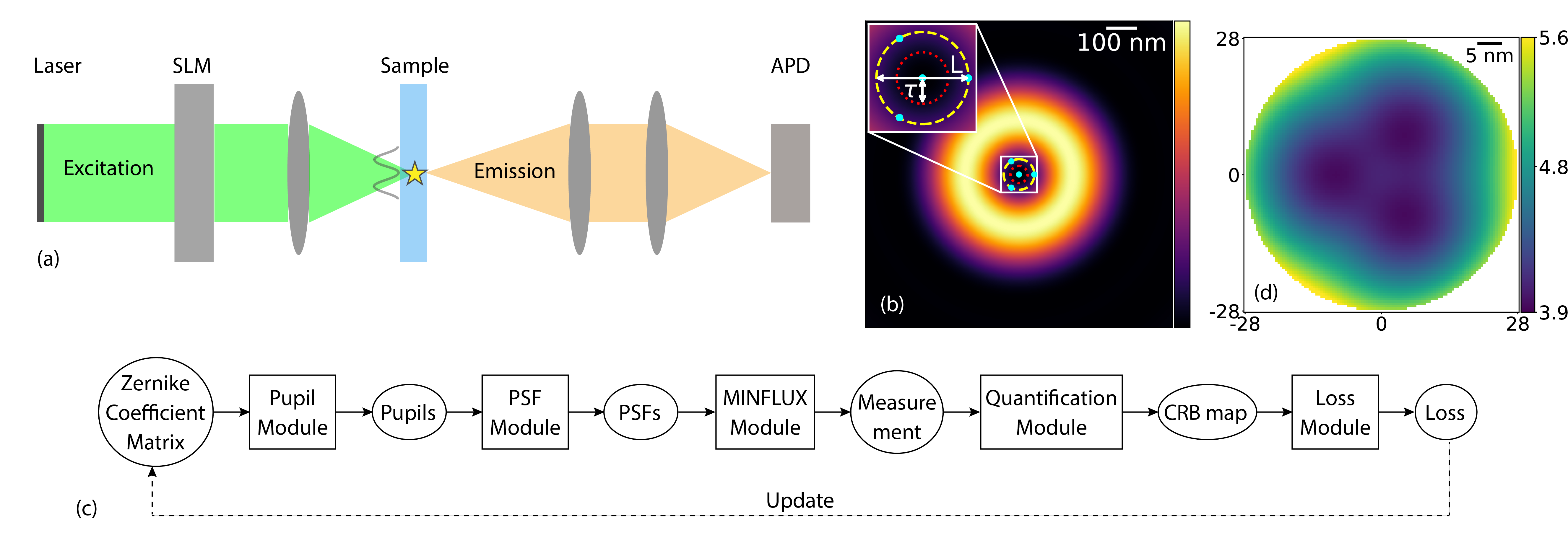}}
    \end{minipage}
    \caption{(a) Diagram of MINFLUX. 
    Excitation and emission light are illustrated in green and orange, respectively. 
    The fluorophore is represented by a yellow star. 
    (b) Intensity distribution of a donut beam on the $xy$-plane.
    The four dots in cyan represent the location of the center of the four donut beams in a standard MINFLUX experiment. 
    Three of these dots lie on a circle with diameter $L=100$ nm (dashed yellow circle).
    The dotted red circle with radius $\tau=28$ nm represents the boundary of the FOV $\Gamma_\tau$ of optimization.
    (c) Pipeline of the optimization problem. 
    Solid and dashed lines with arrowheads indicate the forward and inverse processes, respectively.
    The quantities in the ellipses correspond to the inputs and outputs of the modules in the squares. 
    (d) The CRB map (in nm) in the FOV $\Gamma_\tau$ of a standard MINFLUX experiment with beam setting as in (b) for 100 photons.
    The labels on the horizontal and vertical axes describe the relative distance in nanometers to the center of the FOV for the CRB maps.}
    \label{fig:donut-baseline}
\end{figure*}

The general principle of MINFLUX is as follows. 
The excitation light goes through a beam shaper such as a spatial light modulator (SLM) that modifies the pupil phase such that the resulting beam pattern has a zero of intensity at the focal plane where the sample is placed (Fig. \ref{fig:donut-baseline} (a) and (b)).
Such an isotropic shape has a characteristic low intensity at its center, thus termed the ``zero'' of the beam.
Measurements in MINFLUX correspond to the fluorescence photon counts for different excitation beams, collected by an avalanche photodiode (APD).
The counts are then used to infer the spatial location of the fluorescent target.
In the original MINFLUX experiment \cite{balzarotti_nanometer_2017}, four donut beams are used and arranged into a specific spatial layout: one at the center of the object plane and the other three on a circle of diameter $L$ (Fig. \ref{fig:donut-baseline} (b)) to obtain sufficient information for the inference step.
The smaller $L$ is, the better the localization precision and the smaller the field of view (FOV) where the precision is high. 

We consider the general case of $K$ excitation beams and use a standard Fourier-based model to compute the excitation PSF intensity $h_i(x,y)$ based on the pupil function $P_i$ for $i \in \{ 0, \ldots, K-1\}$
\begin{equation}
    h_i(x,y) = |\F^{-1}\left\{ P_i(k_{\text{x}}, k_{\text{y}}) \right\}(x, y)|^2,
    \label{eq:def-of-PSF}
\end{equation}
where $\mathcal{F}^{-1}$ represents the inverse Fourier transform.
The pupil function $P_i$ is defined as
\begin{equation}
    P_i(k_{\text{x}}, k_{\text{y}}) = \begin{cases}
        \e^{\j \left(\mathbf{c}_i \cdot \mathbf{Z}(k_{\text{x}},\, k_{\text{y}}) + \kappa(k_{\text{x}}, \, k_{\text{y}})\right)}, & k_{\text{x}}^2+k_{\text{y}}^2 \leq k_\text{max}^2 \\
        0, & \text{ elsewhere.}
    \end{cases}
    \label{eq:def-of-phase}
\end{equation}
It is parameterized by the first $D$ Zernike polynomials $\mathbf{Z}(k_{\text{x}}, k_{\text{y}})$ with a coefficient vector $\mathbf{c}_i \in \mathbb{R}^D$, an optional phase term $\kappa(k_{\text{x}},  k_{\text{y}})$ (e.g. a phase ramp for the donut PSF), and a cutoff frequency $k_\text{max} = \frac{2\pi \text{NA}}{\lambda}$ that depends on the numerical aperture $\text{NA}$ and wavelength $\lambda$. 
In response to the $i$th excitation, the number of detected fluorescence photons follows a Poisson process with mean value $\mu_i(x, y) =\alpha h_i(x, y) + b$, where $\alpha$ is a multiplicative constant that depends on the photophysics of the emitter, the detection efficiency, and the energy delivered to the sample, and $b$ is a constant background level including the detector dark counts. Importantly, we add a normalization factor to fix the total expected number of photons across all exposures to $N$. 

To quantify the theoretical localization precision, we first compute the 2-by-2 Fisher information (FI) matrix $I(\bth)$ with respect to the parameter $\bth=(x, y)$ at entry $(j, l)\in\left\{1, 2\right\}$
\begin{equation}
    I(\bth)_{j,l}
    = N\sum_{i=0}^{K-1} 
    \frac{
    \left(\frac{\partial \mu_i}{\partial \theta_j} - \frac{\mu_i\sum_{m=0}^{K-1}\frac{\partial \mu_m}{\partial \theta_j}}{\sum_{m=0}^{K-1}\mu_m}
        \right)
    \left(\frac{\partial \mu_i}{\partial \theta_l} - \frac{\mu_i\sum_{m=0}^{K-1}\frac{\partial \mu_m}{\partial \theta_l}}{\sum_{m=0}^{K-1}\mu_m}
        \right)}
        {\mu_i\sum_{m=0}^{K-1}\mu_m}. 
    \label{eq:def-of-fisher}
\end{equation}
Its derivation is detailed in Section 1.C of the Supplement.
We omitted the spatial dependency of $\mu_i$ for conciseness. 
The CRB is the inverse of the FI matrix, with diagonal elements $\sigma_{\text{x}}^2$ and $\sigma_{\text{y}}^2$, which correspond to a lower bound on the variance of any unbiased estimator of the fluorophore position. 

We define the optimization of the PSFs $h_i$ for $i \in \{ 0, \ldots, K-1\}$, as a minimization problem of the loss function
\begin{equation} 
    J(\mathbf{C}) = \sum_{(x, y)\in \Gamma_{\tau}} \sqrt{\frac{\sigma^2_{\text{x}}(x,y) + \sigma^2_{\text{y}}(x,y)}{2}} - \gamma \sum_{(x, y)\in\Omega_h} \sum_{i=0}^{K-1}h_i(x, y).
    \label{eq:optimization}
\end{equation}
The loss function $J(\mathbf{C})$ is composed of two terms.
The first one is the sum of the averaged CRB values at all the locations within the FOV  of optimization $\Gamma_{\tau}$, a disk of radius $\tau$. 
The second one is the sum of the intensities of all $K$ excitation PSFs within a given region of interest (ROI) $\Omega_\text{h}$. This regularization term encourages that the PSF energy remains inside that region. 
The nonnegative constant $\gamma$ controls the strength of the regularization term. 
This loss function is optimized over the matrix $\mathbf{C}$ of all $K$ Zernike-coefficient vectors $\mathbf{C}=[\mathbf{c}_0, \ldots, \mathbf{c}_{K-1}]$.

The construction of the loss function follows a sequential structure and can be seen as a pipeline composed of modules, as illustrated in Fig. \ref{fig:donut-baseline} (c), with the Zernike-coefficient matrix $\mathbf{C}$ as input and the loss value $J(\mathbf{C})$ as output. 
Leveraging the acyclic nature of the computational graph, we implement it in Pytorch to use its automatic differentiation functionality to obtain the gradient of the loss function, required for the gradient-based optimization of the parameters $\mathbf{C}$.  
The evaluation of \eqref{eq:def-of-fisher} requires an explicit expression of the partial derivative of $\frac{\partial h_i}{\partial\theta_j}$, which is derived in Section 1.E of the Supplement. 
We use a custom fast Fourier transform to establish the optimal discretization parameters for the pupil function and the PSF. 
Based on the chirp Z-transform, it enables us to choose an arbitrarily small PSF pixel size at fixed computational cost. More details on this point and the optimization algorithm are provided in Section 2 of the Supplement. 

We present the results of two distinct groups of experiments aimed at optimizing beam shapes for MINFLUX.
In the first group, we investigate the case in which all beams are translated versions of each other, questioning the optimality of the well-known donut beam. In the second one, we relax the optimization conditions to multiple beam shapes as it has been proposed in recent experiments \cite{Wirth2023, deguchi2024simple}. Key optimization parameters are fixed across both experiments, including numerical aperture $\text{NA} = 1.42$, wavelength $\lambda = 640$ nm, beam arrangement diameter $L = 100$ nm, FOV radius $\tau = 28$ nm for $\Gamma_\tau$, total number of photons $N=100$, and number of Zernike modes $D=15$. 
Despite the inherent complexity of nonlinear optimization, the final beam patterns we present demonstrate remarkable robustness, consistently emerging across a wide range of initial conditions and optimization parameters. 
Following these numerical results, we provide a mathematical analysis that substantiates these beam shapes as optimal in a mathematical sense. 

In the first group, we adopt the original MINFLUX arrangement illustrated in Fig. \ref{fig:donut-baseline} (b), expanded to seven beams to mitigate anisotropy in the CRB map. 
One beam is placed at the center of the field-of-view while the centers of the other six are regularly placed on a circle of diameter $L$~nm.
All seven beams are constrained to have the same shape, a configuration that simplifies experimental implementation as it allows for a fixed phase mask in Fourier space while another optical element (e.g., a galvo mirror) translates the excitation pattern. 

\begin{figure}[tb!]
    \centering
    \includegraphics[width=\columnwidth]{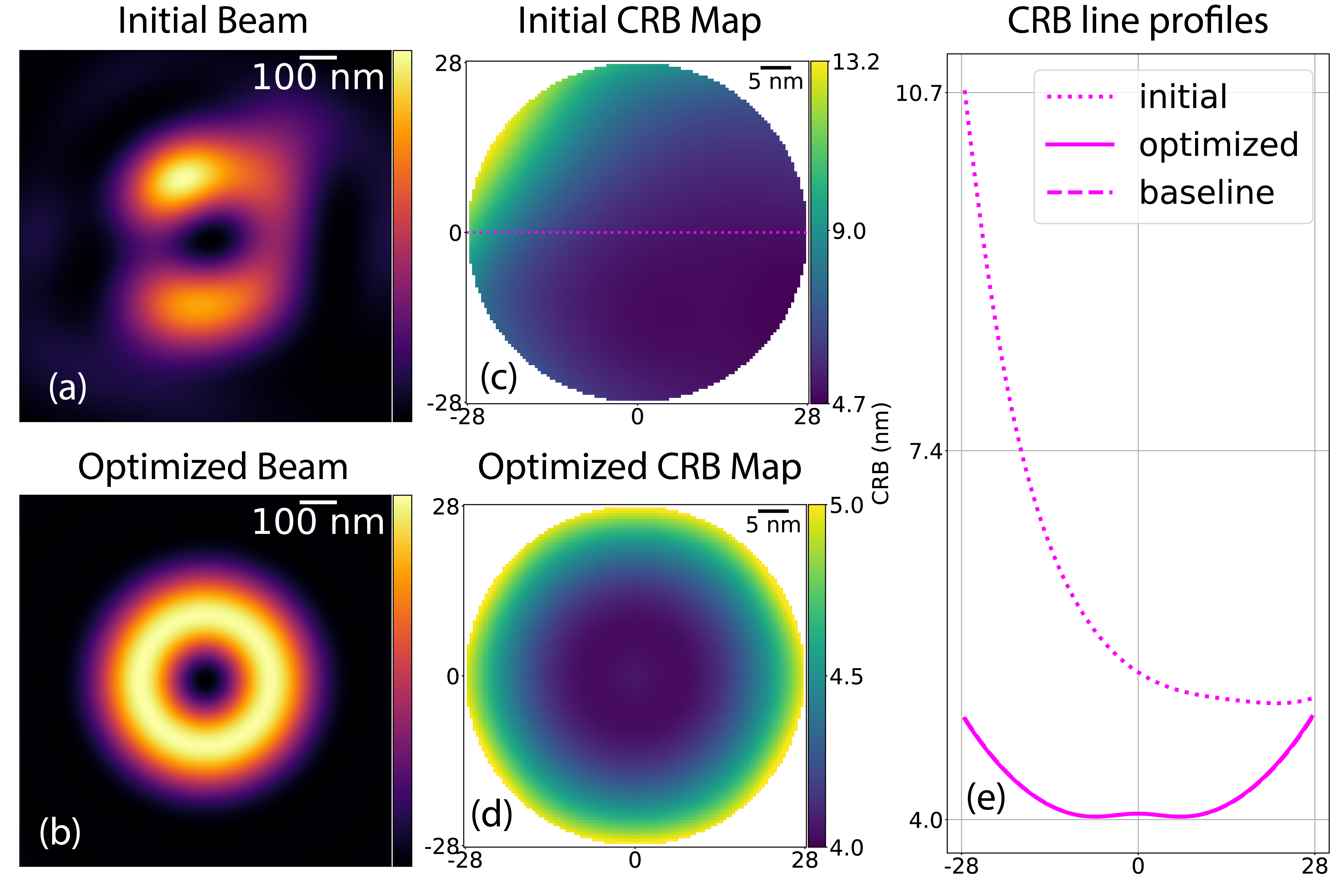}
    \caption{MINFLUX CRB optimization for seven identical excitation beams; one in the center of the FOV and the other six regularly distributed on a circle with diameter $L=100$ nm. 
    (a), (b) initial and optimized intensity of the excitation beam shape;
    (c), (d) initial and optimized CRB map (in nm); 
    (e) comparison of the line profiles in the case of the initial (dotted), optimized (solid) and donut baseline (dashed) CRB maps along the horizontal straight line through the center of the FOV as illustrated in (c).
    }
    \label{fig:one-beam}
\end{figure}

We initiated the optimization with a perturbed donut beam, employing a vortex phase mask $\kappa(k_x, k_y)$ and random initial coefficients $\mathbf{C}$. The results of this optimization, illustrated in Fig.~\ref{fig:one-beam}~(b), exhibit a clear evolution towards a donut shape. Fig. \ref{fig:one-beam} (e) displays the CRB line profiles and reveals a significant decrease in CRB values after optimization. Notably, the CRB values of our optimized beam closely match those of a reference donut beam with seven positions (both curves are superimposed).
To verify the robustness of this result, we conducted additional trials under several initial conditions, including different starting points and levels of background noise. We also directly observed the evolution of the Zernike coefficients. These experiments, presented in the Supplement, consistently converge to the donut shape across all scenarios and firmly establish it as a robust optimal solution for this MINFLUX configuration.

\begin{figure*}[tb!]
    \centering
    \begin{minipage}[b]{0.8\linewidth}
    \centerline{\includegraphics[width=\columnwidth]{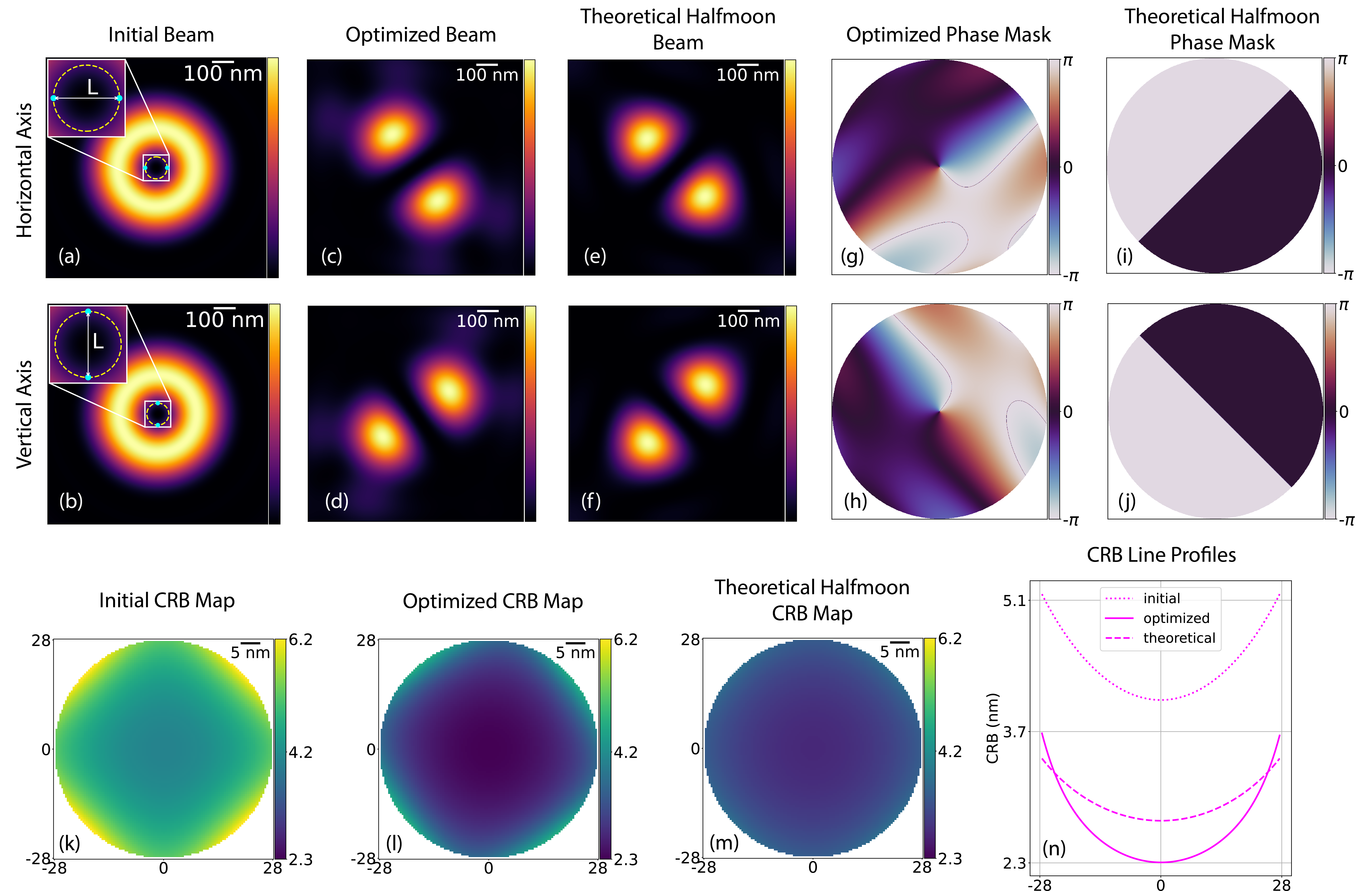}}
    \end{minipage}
    \caption{Diamond setting and two pairs of beam shapes.
    (a), (c), (e) Beams on the horizontal axis for the initial (a), optimized (c), and theoretical (e) cases;
    (b), (d), (f) Beams on the vertical axis for the initial (b), optimized (d), and theoretical (f) cases;
    (g)-(j) Phase masks for the corresponding beams on the horizontal and vertical axes for the optimized and theoretical cases;
    (k)-(m) CRB maps (in nm) of the initial, optimized, and theoretical half-moon cases;
    (n) Horizontal line profiles through the center of the FOV of the CRB maps (k)-(m).
    }
    \label{fig:two-pairs}
\end{figure*}

In our second group of experiments, we introduced a configuration using two distinct pairs of beams. We arranged these beams in a square pattern, using four donut beams as our starting configuration. The center of these four beams are arranged on a circle of diameter $L=100$ nm at locations $(\frac{L}{2}, 0), (0, \frac{L}{2}), (-\frac{L}{2}, 0)$ and $(0, -\frac{L}{2})$ in x-y coordinates as depicted in Fig. \ref{fig:two-pairs} (a) and (b). 

The PSF shapes evolved significantly during optimization and converged to a configuration characterized by two approximately symmetric peaks separated by a valley of low intensities (Fig. \ref{fig:two-pairs} (c) and (d)). 
Further, we notice that the axes of the two pairs of beams are orthogonal to each other, oriented along the edges of the square formed by the beam centers. 
The associated phase masks of these beams are shown in Fig. \ref{fig:two-pairs} (g) and (h). 
This optimized configuration significantly improves the CRB values by a factor of two (Fig. \ref{fig:two-pairs} (k) and (l)). 

The optimal beams resemble the existing ``half-moon'' beams which have been seen in \cite{Wirth2023}. In practice, these shapes can be generated by a $\pi$-phase-shifted half-disk in the pupil plane (Fig. \ref{fig:two-pairs} (i) and (j)).
These theoretical half-moon beams yield slightly more uniform CRB maps compared to the optimized PSFs, as shown in Fig. \ref{fig:two-pairs} (l) and (m). 
A comparison of the CRB line profiles for the initial, optimized,  and theoretical half-moon beams is shown in Fig. \ref{fig:two-pairs} (n).
We tested the robustness of the half-moon shape across various optimization conditions, including scenarios where all four beams were allowed to have independent shapes. These results are provided in the Supplement. Consistently, the half-moon configuration emerged as a stable and optimal solution.

To complement our optimization results, we discuss theoretical results in Theorem 1 and Proposition 2.
We demonstrate how the donut and half-moon PSFs are optimal in a specific mathematical sense. 
A more detailed analysis, along with a rigorous proof of the theorem and numerical results, is presented in Section 3 in the Supplement. 
Precision in MINFLUX is linked to the steepness of the intensity variations near the center of the region of interest \cite{balzarotti_nanometer_2017, Wirth2023}, which leads to our study on how to maximize the norm of the gradient of the electric field. 

\textbf{Theorem 1}
\emph{The half-moon beam maximizes the directional derivative norm of the electric field at the center of the field of view.}

\textbf{Proposition 2}
\emph{The donut beam maximizes the angular average of the squared norm of the directional derivative of the electric field.}

Theorem 1 accounts for the half-moon shape obtained in our second group of experiments, while Proposition 2 provides insights into the optimality of the donut shape when all beams are constrained to have the same shape. 
We obtain this last result by formulating it as a quadratic programming problem with modulus constraint and finding the numeric optimum using projected power iterations. 

Our theoretical results establish that the beam shapes obtained from our optimization framework maximize the aforementioned metrics, representing fundamental optima. 
However, they do not account for shifts and the estimation problem, which necessitates our Fisher-information framework.

To conclude, our study leads to two key findings in the PSF engineering of MINFLUX.
The donut beam is optimal when the excitation pattern is constrained to take a single shape, while the half-moon beams yield lower CRB when multiple shapes are available.
These results, obtained through our PSF-engineering pipeline, are further supported by a mathematical investigation of the maximization of the gradient of the electric field.
Our theoretical framework is a fast and economic approach to discover and investigate novel beam shapes for MINFLUX.
Additionally, this flexible framework can be extended to optimize PSFs for 3D localization using vectorial models of the electric field propagation, potentially expanding the capabilities of MINFLUX across diverse imaging scenarios.

\begin{backmatter}
\bmsection{Funding} This project was supported by the Swiss National Science Foundation (SNSF) under Grant PZ00P2\_216211 and the European Research Council (ERC) under the European Union’s Horizon 2020 research and innovation program (Grant agreement No. 853348). F.B. was supported by Boehringer Ingelheim. 

\bmsection{Acknowledgments} We would like to thank Eric Sinner for his valuable help in the code base and Mehrta Shirzadian for fruitful discussions.

\bmsection{Disclosures} F.B. holds patents on principles, embodiments and procedures of MINFLUX. The other authors declare no conflicts of interest.

\bmsection{Data availability} 
The code to produce the results in this letter is available online\footnote{\href{https://zenodo.org/records/13857972}{https://zenodo.org/records/13857972}}.

\end{backmatter}

\bibliography{bib.bib}

\newpage
\onecolumn

\appendix

\section{Theoretical Foundation: Physical and Mathematical Models}\label{sec:theory}
\subsection{Scalar Excitation PSF model via Zernike Polynomials}\label{sec:psf-engineering}

We use a scalar model \eqref{eq:def-of-PSF}for the beam propagation.
It describes the relation between the pupil function $P$ in the Fourier space and the resulting PSF $h$ at the focal plane in the object space.
The wave vector in the Fourier plane is $(k_\text{x},k_\text{y})$.
The pupil function $P: \mathbb{R}^2 \to \mathbb{C}$ is further defined as \cite{shechtman_optimal_2014}
\begin{equation}
    P(k_\text{x}, k_\text{y}) = \mathrm{circ}\left(\frac{\sqrt{k_\text{x}^2 + k_\text{y}^2}}{k_{\max}}\right)\exp\left\{\j\left(\sum_{d=0}^{D-1} c_d \varphi_d(k_\text{x}, k_\text{y}) + \kappa(k_\text{x}, k_\text{y}) \right)\right\},
    \label{eq:def-of-pupil}
\end{equation}
with $\j$ the imaginary unit and $k_{\max}=\frac{2\pi NA}{\lambda}$ the cutoff in the Fourier space.
The function circ$(\cdot)$ defines the unit disk
\begin{equation}\label{eq:circ}
    \mathrm{circ}(r) = \left\{
    \begin{array}{cc}
       1,  &   r<1,\\
       0,  &   r\geq 1.
    \end{array}
    \right.
\end{equation} 
The series of Zernike coefficients is represented by a $D$-dimensional vector $\mathbf{c} = (c_0, \ldots, c_d, \ldots, c_{D-1} )$, where $c_d\in \mathbb{R}$, $\left\{ \varphi_d(\cdot, \cdot) \right\}_{d=0}^{D-1}$ are the  first $D$ Zernike polynomials and  $\varphi_d:\mathbb{R}^2 \to \mathbb{R}$. 
An extra phase factor $\kappa(k_\text{x}, k_\text{y})$ is added to the phase. 
The complex term in the exponential in \eqref{eq:def-of-pupil} represents a phase mask that modulates the shape of the resulting excitation PSF. 
When the phase mask is the ramp $\kappa(k_\text{x},k_\text{y}) = \text{arctan}\left(\frac{k_\text{x}}{k_\text{y}}\right)$, the resulting PSF is a donut beam.

\subsection{Statistical Model of the Photon Detection in MINFLUX}
In MINFLUX, an emitter (e.g., a fluorescent molecule) located at $\bth=(x, y)$ is exposed to a total number of $K$ different excitation beams.
We describe the intensities of these beams by the following $K$ functions $h_0(\bth), \ldots, h_{K-1}(\bth)$, where $h_i(\bth) = h(\bth - \t_i)$, with $\t_i = (t_{\text{x}, i}, t_{\text{y}, i})\in\R^2$ constant shift vector. We define the expected number of detected photons for the $i$-th exposure before normalization $\mu_i(\bth)$ as
\begin{equation}\label{eq:def-mu}
    \mu_i(\bth) = \alpha h_i(\bth) + b.
\end{equation}
The constant $\alpha$ depends on the photo-physics of the emitter, the detection efficiency of the system, and the energy delivered to the sample, and $b$ is a background term. 

We introduce a position-dependent normalization factor $\N(\bth)$ to impose a budget $N$ on the total expected number of photons on the detector for any position $\bth$ to ensure a fair comparison of the detection limits between experiments. 
The expected number of detected photons after normalization is 
\begin{equation}
    n_i(\bth) = \N(\bth) \mu_i(\bth),
\end{equation}
and the normalization factor is defined as
\begin{equation}
    \N(\bth) = \frac{N}{\sum_{m=0}^{K-1} \mu_i(\bth)}
    \label{eq: definition normalization}
\end{equation}
such that $N = \sum_{m=0}^{K-1} n_i(\bth)$.

We model the number of detected photons for the $i$-th exposure as a Poissonian random variable (R.V.) $X_i$ with mean $n_i(\bth)$ dependent on the emitter position $\bth\in\Gamma_\tau\subset\mathbb{R}^2$. Its probability mass function (PMF) is
\begin{equation}\label{eq:poisson-pdf}
    f_i(s_i; \bth) = \P(X_i=s_i; \bth) = \frac{n_i(\bth)^{s_i}\e^{-n_i(\bth)}}{s_i!}, \quad i=0, \ldots, K-1,
\end{equation}
Finally, the measurements of a MINFLUX experiment is a list of photon counts $s_0,\dots, s_{K-1}$ which are realizations of the $K$ R.V.'s $(X_0, \ldots, X_{K-1})$. 
The photons emitted by the fluorescent molecule and the background are statistically independent. 

\subsection{Statistical Detection Limit via the Cram\'er-Rao Lower Bound}\label{sec:fisher-CRB}
We would like to estimate the multivariate parameter $\bth$ using $K$ independent measurements $(X_0, \ldots, X_{K-1})$. The Fisher information for estimating $\bth$ then is defined as
\begin{equation}\label{eq:def-fisher}
    I(\bth)_{j, l} = \sum_{i=0}^{K-1} -\E\left[\frac{\partial^2 \log (f_i(X_i; \bth))}{\partial\theta_l\partial\theta_j}\right],
\end{equation}
where the expectation $\E=\E_{X_i;\bth}$ is taken over the R.V. $X_i$, the indices $j, l\in\left\{1, 2\right\}$.
Now we proceed to compute \eqref{eq:def-fisher} step by step.
First compute $\log(f_i(X_i; \bth))$ as
\begin{equation}\label{eq:log-f}
    \log(f_i(X_i; \bth)) = X_i\log n_i(\bth) - n_i(\bth) - \log (X_i!)
\end{equation}
Then, its first- and second-order partial derivatives are
\begin{equation}\label{eq:partial-log-f}
    \frac{\partial\log(f_i(X_i; \bth))}{\partial\theta_j} = X_i\frac{1}{n_i(\bth)}\frac{\partial n_i(\bth)}{\partial \theta_j} - \frac{\partial n_i(\bth)}{\partial\theta_j}
\end{equation}
\begin{equation}\label{eq:partial2-log-f}
    \frac{\partial^2\log(f_i(X_i; \bth))}{\partial\theta_l\partial\theta_j} = X_i\left[ 
    -\frac{1}{n_i(\bth)^2} \frac{\partial n_i(\bth)}{\partial \theta_l} \frac{\partial n_i(\bth)}{\partial \theta_j}
    + \frac{1}{n_i(\bth)} \frac{\partial^2 n_i(\bth)}{\partial \theta_l \partial \theta_j}
    \right]
    - \frac{\partial^2 n_i(\bth)}{\partial \theta_l \partial \theta_j}.
\end{equation}
Finally, take the expectation to attain that
\begin{align}\nonumber
    \E\left[
    \frac{\partial^2\log(f_i(X_i; \bth))}{\partial\theta_l\partial\theta_j}
    \right] & = n_i(\bth)\left[ 
    -\frac{1}{n_i(\bth)^2} \frac{\partial n_i(\bth)}{\partial \theta_l} \frac{\partial n_i(\bth)}{\partial \theta_j}
    + \frac{1}{n_i(\bth)} \frac{\partial^2 n_i(\bth)}{\partial \theta_l \partial \theta_j}
    \right]
    - \frac{\partial^2 n_i(\bth)}{\partial \theta_l \partial \theta_j} \\ \nonumber
    & = -\frac{1}{n_i(\bth)}\frac{\partial n_i(\bth)}{\partial \theta_l} \frac{\partial n_i(\bth)}{\partial \theta_j}.\label{eq:exp}
\end{align}
Hence,
\begin{equation}\label{eq:fisher-expression}
    I(\bth)_{j, l} = \sum_{i=0}^{K-1}\frac{1}{n_i(\bth)}\frac{\partial n_i(\bth)}{\partial \theta_l} \frac{\partial n_i(\bth)}{\partial \theta_j} = \sum_{i=0}^{K-1} 
    \frac{1}{\N (\bth)  \mu_i  (\bth)} \frac{\partial (\N (\bth)  \mu_i  (\bth))}{\partial \theta_l} \frac{\partial(\N (\bth)  \mu_i  (\bth))}{\partial \theta_j}.
\end{equation}

Now, we proceed to simplify this last expression.
First, we handle the partial derivative term by plugging \eqref{eq: definition normalization} in $\frac{\partial (\N (\bth)  \mu_i  (\bth)))}{\partial \theta}$ which leads to
\begin{align}\nonumber
    \frac{\partial (\N(\bth)  \mu_i(\bth) )}{\partial \theta} =& 
    \frac{-N \sum_{m=0}^{K-1}\frac{\partial \mu_m}{\partial \theta} }{\left(\sum_{m=0}^{K-1}  \mu_m(\bth)\right)^2} \mu_i(\bth) 
  + \frac{N}{\sum_{m=0}^{K-1}  \mu_m(\bth)}\frac{\partial  \mu_i(\bth) }{\partial \theta} \\
    =& \frac{N}{\sum_{m=0}^{K-1}  \mu_m(\bth)}\left(\frac{\partial  \mu_i(\bth) }{\partial \theta} - \frac{\sum_{m=0}^{K-1}\frac{\partial \mu_m}{\partial \theta} }{\sum_{m=0}^{K-1}  \mu_m(\bth)}\mu_i(\bth)\right).\label{eq:derivative-modified}
\end{align}

Finally, we obtain the final Fisher Information matrix expression
\begin{equation}\label{eq:fisher-final}
    I(\bth)_{j,l}
    = \sum_{i=0}^{K-1} 
    \frac{N}{ \mu_i(\bth) \sum_{m=0}^{K-1} \mu_m(\bth)}
    \left(\frac{\partial  \mu_i(\bth)}{\partial \theta_j} - \frac{ \sum_{m=0}^{K-1}\frac{\partial \mu_m}{\partial \theta_j}}{\sum_{m=0}^{K-1} \mu_m(\bth)} \mu_i(\bth)
        \right)
    \left(\frac{\partial  \mu_i(\bth)}{\partial \theta_l} - \frac{\sum_{m=0}^{K-1}\frac{\partial \mu_m}{\partial \theta_l} }{\sum_{m=0}^{K-1} \mu_m(\bth)} \mu_i(\bth)
        \right). 
\end{equation}
The statistical detection limit is determined by the Cram\'er-Rao lower bound.
It indicates the theoretical best detection precision of MINFLUX. Said otherwise, \eqref{eq:fisher-final} indicates the minimal variance of any unbiased estimator of $\bth$. 
Assuming the invertibility of the Fisher information matrix $I$, we denote $I^{-1}$ its inverse and
\begin{equation}
I^{-1}=
    \begin{bmatrix}
    \sigma^2_\text{xx} & \sigma^2_\text{xy} \\
    \sigma^2_\text{yx} & \sigma^2_\text{yy}
    \end{bmatrix},
\end{equation}
where the diagonal entries $\sigma^2_\text{xx}$ and $\sigma^2_\text{yy}$ are the variance of the estimated $x$ and $y$ locations, respectively.

\subsection{Explicit Expression of the Fisher Information Matrix}\label{sec:partial-derivatives}

In \eqref{eq:fisher-final} we need to evaluate the partial derivatives of $\mu_i$ at location $(x - t_x, y - t_y)$ with respect to $x$ or $y$. 
Here we provide an analytical expression for these partial derivatives. 

Denote the inverse Fourier transform of the pupil function as
\begin{equation}
    g(x, y) = \F^{-1}\left\{P(k_x, k_y)\right\}(x,y)
\end{equation}
$h$ is defined as the square modulus of $g$ which can be written as the element-wise multiplication of $g$ and its complex conjugate $\g$: $h(x, y) = |g(x, y)|^2 = g(x, y)\g(x, y)$.

The derivative of $h$ with respect to $x$ at location $(x - t_x, y - t_y)$ is
\begin{eqnarray}
    \frac{\partial h}{\partial x}(x - t_x, y - t_y) &=& \frac{\partial (g(x - t_x, y - t_y) \g(x - t_x, y - t_y) )}{\partial x}\\\nonumber
    &=& \frac{\partial g}{\partial x}(x - t_x, y - t_y) \g(x - t_x, y - t_y) \\\nonumber
    && + \frac{\partial \g}{\partial x}(x - t_x, y - t_y) g(x - t_x, y - t_y)
    \\
    &=& 2 \cdot \text{Re}\left( \frac{\partial g}{\partial x}(x - t_x, y - t_y) \g(x - t_x, y - t_y)\right)\label{eq:h-derivative}.
\end{eqnarray}

Using the shift property of the Fourier transform,
\begin{equation}
    \g(x - t_x, y - t_y) = \overline{\F^{-1}\left\{\e^{-\j(k_x t_x + k_y t_y)} P(k_x, k_y)\right\}(x, y)}.
    \label{eq:p-bar-shifted}
\end{equation}
Using the differentiation and shift property of the Fourier transform, 
\begin{equation}
    \frac{\partial g}{\partial x}(x - t_x, y - t_y) = \F^{-1} \left\{ (\j k_x)\e^{-\j (k_x t_x + k_y t_y)} P(k_x, k_y) \right\}(x, y).
    \label{eq:dpdx}
\end{equation}

Combining \eqref{eq:dpdx} and \eqref{eq:p-bar-shifted}, we have an analytical expression for $\frac{\partial h}{\partial x}(x - t_x, y - t_y)$. Analogously, the derivative of $h$ with respect to $y$ at location $(x - t_x, y - t_y)$, $\frac{\partial h}{\partial y}(x - t_x, y - t_y)$,  can be obtained by replacing $\j k_x$ with $\j k_y$ in the above expression. Finally, the partial derivatives of $\mu_i$ are easily derived using \eqref{eq:def-mu}.

\subsection{Definition of the Optimization Problem}
We are interested in the optimal shape of the excitation beams and look for a set of Zernike coefficients that would lead to a minimal CRB under certain experimental setups.
We define such a quest as the following minimization problem of a loss function $J$ defined in \eqref{eq:optimization}. 
The columns in the $(D\times K)$ matrix $\mathbf{C} = [\mathbf{c}_0, \ldots, \mathbf{c}_{K-1}]$ contain the Zernike coefficient vectors of all $K$ beams.
The loss is composed of two terms.
The relation of $\mathbf{c}$ and $h_i$ is provided in \eqref{eq:def-of-PSF} and \eqref{eq:def-of-pupil}.
The first term in $J(\mathbf{C})$ is a sum of  $\nu(x, y)$, defined as the arithmetic average of the variance terms $\sigma_\text{xx}^2 $ and $ \sigma_\text{yy}^2$ 
\begin{equation}
    \nu(x, y) = \sqrt{\frac{\sigma_\text{xx}^2 + \sigma_\text{yy}^2}{2}}.
\end{equation}
of every location $(x, y)$ inside a given field of view $\Gamma\subset\R^2$.
It is the primary quantity that we aim to minimize.
The second term involves a nonnegative constant $\gamma$ multiplied by the sum of the intensities of all $K$ beams inside a given region of interest $\Omega_\text{h}\subset\mathbb{R}^2$.
The hyper-parameter $\gamma$ controls the strength of regularization that helps us guide the optimization into directions of desirable beam shapes. 
To be precise, here it is defined to maximize the intensity of all the beams inside the ROI and avoid degenerate cases where most of the energy of the beams is outside the FOV.
Because the problem \eqref{eq:optimization} is nonconvex with multiple local minima, the solution is not unique and strongly depends on the initial start.

In our simulations, we define $\Gamma$ as a disk the radius of which is a tunable hyperparameter. 
The ROI $\Omega_\text{h}$ is defined similarly as a disk with a radius of 170 nm, approximately the radius of a donut beam.

\clearpage
\section{Implementation of the optimization pipeline}
In this section, we present the optimization algorithm and an important technical detail, the chirp Z-transform which allows us to efficiently compute the inverse Fourier transform.

\subsection{Optimization Algorithm}\label{sec:optimization}
We provide the following Algorithm S1 to solve \eqref{eq:optimization}.
\begin{algorithm}[ht]
\caption{Optimization algorithm}\label{alg:optimization}
\begin{algorithmic}[1]
\Require Initial guess of the Zernike coefficient matrix $\mathbf{C}^0$, stopping threshold $\epsilon=10^{-6}$ 
\State $N_{\textrm{iter}} = 0$
\State Calculate the loss $J(\mathbf{C}^{N_{\textrm{iter}}})$ in \eqref{eq:optimization}
\State Calculate the gradient $\frac{\mathrm{d}J}{\mathrm{d}\mathbf{C}} (\mathbf{C}^{N_{\textrm{iter}}})$ using \texttt{J.backward()}
\State Gradient-descent step with built-in optimizer: \texttt{Adam(lr=0.05, betas=(0.9, 0.999))}
\If {$N_{\textrm{iter}} > 0 $}
    \If {$|J(\mathbf{C}^{N_{\mathrm{iter}}}) - J(\mathbf{C}^{N_{\textrm{iter} - 1}})| > \epsilon$} 
        \State Repeat Steps 2 to 4
        \State $N_{\textrm{iter}}\leftarrow N_{\textrm{iter}} + 1$ 
    \Else 
        \State Terminate
    \EndIf
\EndIf
\end{algorithmic}
\end{algorithm}

The coefficient matrix $\mathbf{C}^{N_{\textrm{iter}}}$ of the final iteration is used to generate the final optimized pupil, the PSF and the corresponding CRB map. 

All results share the numerical hyperparameters. The size of the object plane is $(1000 \times 1000)$ nm. The background level $b$ is set to be $5\%$ of the mean value of $h$ on the whole object plane. The total number of photons received on the detector is $N=100$. The efficiency $\alpha$ of the experiment is assumed to be 1 and the beam positions are defined with $L = 100$ nm. The radius $\tau$ of the optimization FOV is 28 nm, the regularization term is computed on a ROI $\Omega_\text{h}$ of radius 170 nm, approximately the radius of the donut beam. The regularization weight $\gamma$ is 0.1.

Our pipeline has a practical multi-resolution feature. 
Because the input of the pipeline is a Zernike coefficient matrix that depends neither on the numerical size of the pupil nor on the PSF, we are able to run the optimization on a coarse meshgrid to save computational time and generate high-resolution images using the optimized coefficients.
For optimization, the numerical sizes of the pupil and PSF are $(256\times256)$ px and $(512\times512)$ px, respectively.
This increases computational speed while retaining sufficient sampling of the pupil function and the CRB map.
After optimization, we run an extra round of the full forward pipeline that takes optimized Zernike coefficients and a larger number of sampling points.  
All the images in both the manuscript and the Supplement are of size $(2048\times2048)$ px unless specified otherwise.

The computational time and the evolution of the loss of each optimization experiment presented in the manuscript and the supplement are shown in Fig. \ref{fig:loss}.

\subsection{Efficient Computation of the Generalized Inverse Fourier Transform}\label{sec:z-tranaform}
We see from Sections \ref{sec:psf-engineering} and \ref{sec:partial-derivatives} that the inverse Fourier transform (IFT) is frequently required to compute the beam-propagation model and the partial derivatives found in the Fisher matrix.
There are two notable aspects to consider in our specific case.
First, the sizes of the input and output of the IFT can be different.
Moreover, we would like to zoom in onto the object plane to refine PSF function (partial sampling).
The ordinary IDFT is unfortunately unable to satisfy either aspect.
In this section, we address the shortcomings of the ordinary IDFT by using the chirp Z-transform (CZT) and provide an efficient algorithm via the CZT.
It has the same computational complexity as the FFT.

\begin{figure}[ht]
    \centering
    \includegraphics[width=0.6\textwidth]{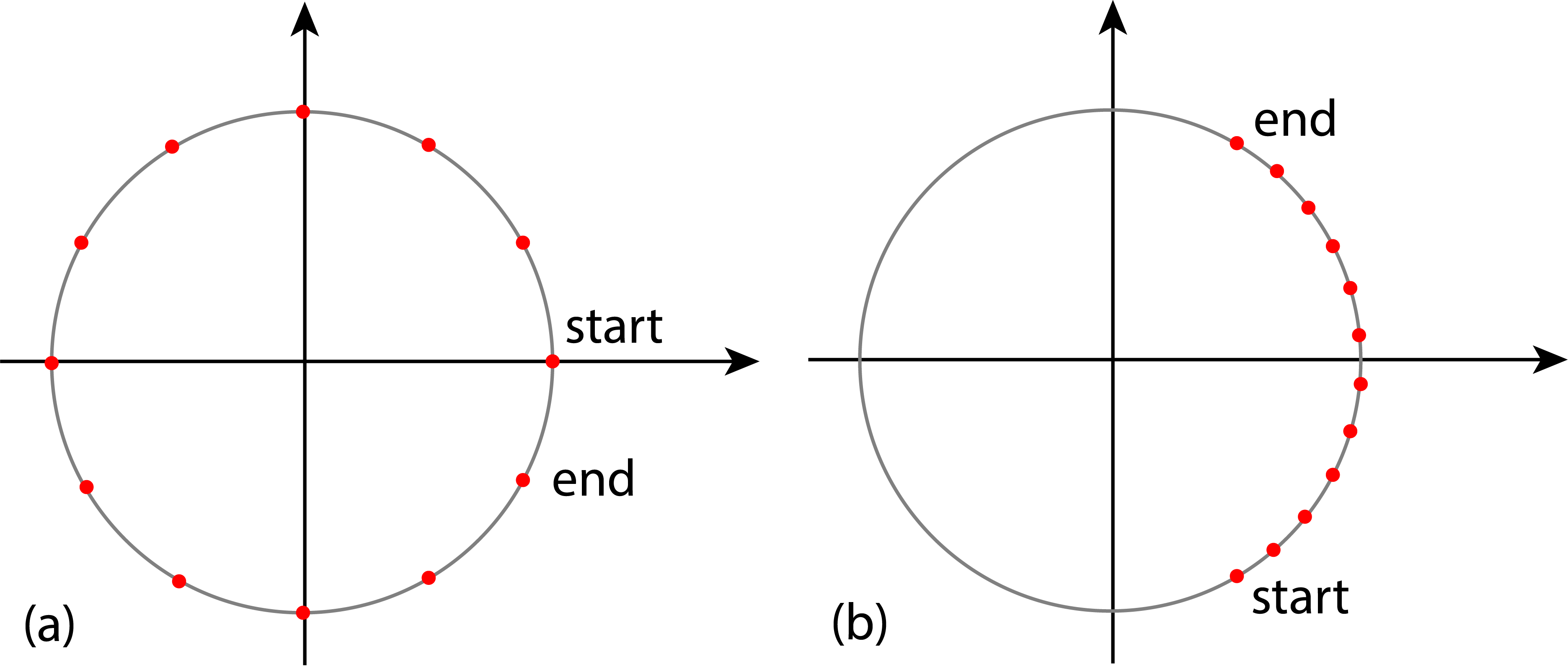}
    \caption{Sampling of $N$ points on the unit circle. (a) IDFT samples uniformly the whole unit circle. (b) zoomIDFT performs a uniform sampling of a subset of the unit circle.}
    \label{fig:fft-sampling}
\end{figure}

The chirp Z-transform of a discrete 1D signal $x_n$ of length $N$ is defined as 
\begin{equation}\label{eq:z-transform}
    y_m = \sum_{n=0}^{N-1} x_n a^{-n} w^{nm}, \quad m=0,\ldots, M-1,
\end{equation}
where $y_m$ is the $m$th sample of the output signal of length $M$, $M$ being not necessarily equal to $N$.
We observe that IDFT is a special case of \eqref{eq:z-transform} with $a=1$ and $w=\e^{\frac{\j}{N}}$ and a scaling factor of $\frac{1}{N}$.
To be precise, the sampling of $m$ starts from $(1, 0)$ and covers the whole unit circle with a constant angular step size of $\frac{2\pi}{N}$ on the complex plane, see Fig. \ref{fig:fft-sampling}(a).
We define a special IFFT, termed zoomIDFT
\begin{equation}\label{eq:zoomifft}
    y^z_m = \textrm{zoomIDFT}(x_n) \coloneqq  \frac{1}{N} \sum_{n=0}^{N-1} x_n \e^{\frac{\j}{N} z (n - \frac{N}{2}) (m-\frac{M}{2})}, \quad m=0,\ldots, M-1,
\end{equation}
where $z\in(0, 1]$ is the zoom factor such that only a fraction $z$ of the full unit circle is sampled.
Further, the origin of both the input and output signals is centered as indicated by $(n - \frac{N}{2})$ and $(k-\frac{M}{2})$, see Fig. \ref{fig:fft-sampling}(b).

\begin{figure}[ht]
    \centering
    \includegraphics[width=0.9\textwidth]{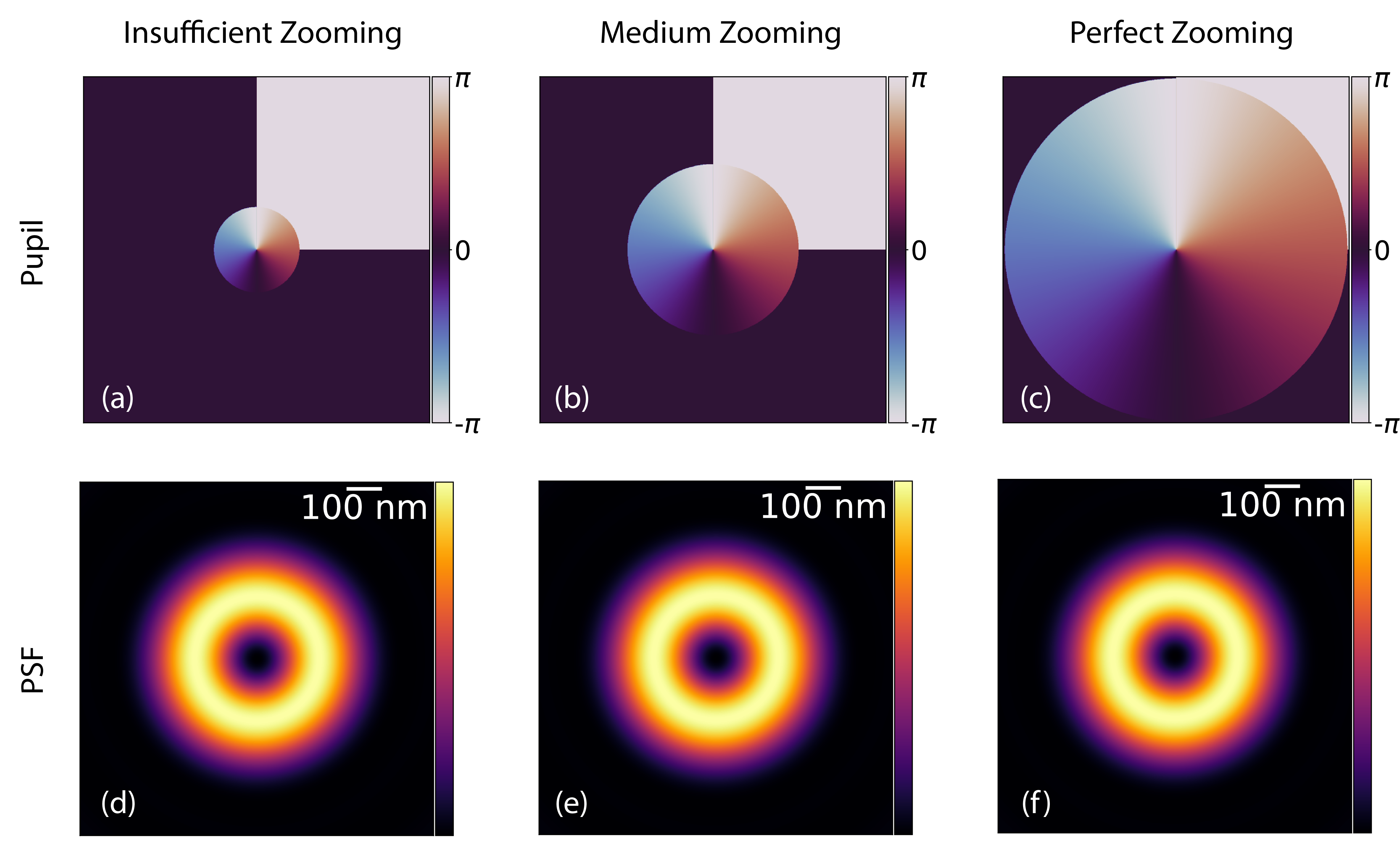}
    \caption{Zooming factor. 
    First row: Pupil functions with an insufficient (a) zooming factor of 0.00875, medium (b) factor of 0.004375, and perfect (c) zooming of 0.0021875.
    Second row: Corresponding PSFs.
    The numerical sizes of the pupils and PSFs are $(1028\times1028)$ px and $(2048\times2048)$ px, respectively. }
    \label{fig:zooming}
\end{figure}

Given the desired number of sampling points for both the Fourier and object planes, zoomIDFT allows us to control the area on which the pupil occupies via a constant --- the ``zooming factor'' (See Fig. \ref{fig:zooming}).
Ideally, the zooming factor is chosen such that the pupil fully occupies the whole plane for maximal numerical efficiency.

To efficiently compute \eqref{eq:zoomifft}, we follow the techniques proposed in \cite{Bluestein1970, Rabiner1969, Leutenegger:06} and rewrite \eqref{eq:z-transform} as the convolution
\begin{eqnarray}\label{eq:conv}
    y_m &=& w^{\frac{m^2}{2}}\sum_{n=0}^{N-1}\left(x_n a^{-n} w^{\frac{n^2}{2}} \cdot w^{-\frac{(m-n)^2}{2}}\right)\\
    &=&w^{\frac{m^2}{2}} \left((x_n a^{-n} w^{\frac{n^2}{2}}) * w^{-\frac{n^2}{2}}\right) \\\label{eq:ffts}
    &=& w^{\frac{m^2}{2}} \cdot \textrm{IFFT}\left( \textrm{FFT}(x_n a^{-n} w^{\frac{n^2}{2}}) \cdot \textrm{FFT}(w^{-\frac{n^2}{2}})\right)
\end{eqnarray}
using the relation $nm = (n^2 + m^2 - (m-n)^2) / 2$ \cite{Bluestein1970} to separate the variables $n$ and $m$.
\eqref{eq:conv} is then efficiently computed using two precomputed FFTs and one IFFT. 
We thus shuffle the order of the terms of zoomIFFT in \eqref{eq:zoomifft} to obtain that
\begin{equation}\label{eq:zoomifft-compute}
    y^z_m = \e^{-\frac{\j}{2} z(m-\frac{M}{2})} \cdot \frac{1}{N}\sum_{n=0}^{N-1} x_n \e^{(\frac{\j}{2} z\frac{M}{N})(-n)} \e^{(\frac{\j}{N}z) (nm)}, 
\end{equation}
where $a = \e^{\j z\frac{M}{N}}$, $w=\e^{\frac{\j}{N}z}$.
Then, we use \eqref{eq:ffts} to compute the CZT contained in \eqref{eq:zoomifft-compute}.
\newpage

\clearpage
\section{Mathematical Analysis of the Optimality of the Donut and Half-Moon Beams}
\subsection{Context}
This section is dedicated to a mathematical analysis to further understand the optimality of the half-moon and donut PSFs obtained through our CRB optimization. We demonstrate that these PSFs maximize the norm of the gradient of the electric field at the center of the field of view: for a single direction in the half-moon case, and averaged over all directions for the donut PSF. The half-moon case relies on an analytic proof, while the donut case relies on numerical evidence. These results provide a heuristic explanation for the final PSF forms; they do not account for shifts and the estimation problem, which necessitates our Fisher-information framework. 

To ease the explanation, let us recall the definition of the PSF intensity $h: \mathbb{R}^2 \rightarrow \mathbb{C}$
\begin{equation}
    \label{eq: intensity}
    h(\r) = |g(\r)|^2,
\end{equation}
where $\r = (x,y)$ and the electric field $g: \mathbb{R}^2 \rightarrow \mathbb{C}$ is given by
\begin{equation}
    \label{eq: electric field}
    g(\r) = \F^{-1}\left\{ P(\k) \right\}(\r).
\end{equation}
The pupil function $P(\k): \Lambda \rightarrow \mathbb{C}$ is defined on a finite support $\Lambda \subset \mathbb{R}^2$, given by the low-pass-filtering disk that corresponds to the numerical aperture, with the constraint that $P$ has unit modulus over $\Lambda$ so that
\begin{equation}
    \label{eq: pupil constraints}
    \begin{cases}
        |P(\k)| = 1, \qquad & \k \in \Lambda,\\
        P(\k) = 0, \qquad &\text{else}.
    \end{cases}
\end{equation}

The problem is two-dimensional; yet, we start by explaining our heuristic in 1D. MINFLUX minimizes the photon flux by precisely targeting and modulating the emission of fluorophores near the zero region of the excitation PSF. 
We thus assume in 1D that $h(0) = g(0) = 0$ and express the Taylor expansion of $g$ as $g(x) = g'(0) x + \mathcal{O}(x^2)$, which implies that the approximated PSF $\tilde{h}(x) = |g'(0)|^2 x^2$ is a parabola that is centered. We aim to maximize the curvature coefficient $|g'(0)|^2$ to maximize the excitation change for a change in the particle position. This heuristic was first introduced in the original MINFLUX paper \cite{balzarotti_nanometer_2017}.

\subsection{Optimization for a Single Direction}
\begin{figure}
    \centering
    \includegraphics[width=\textwidth]{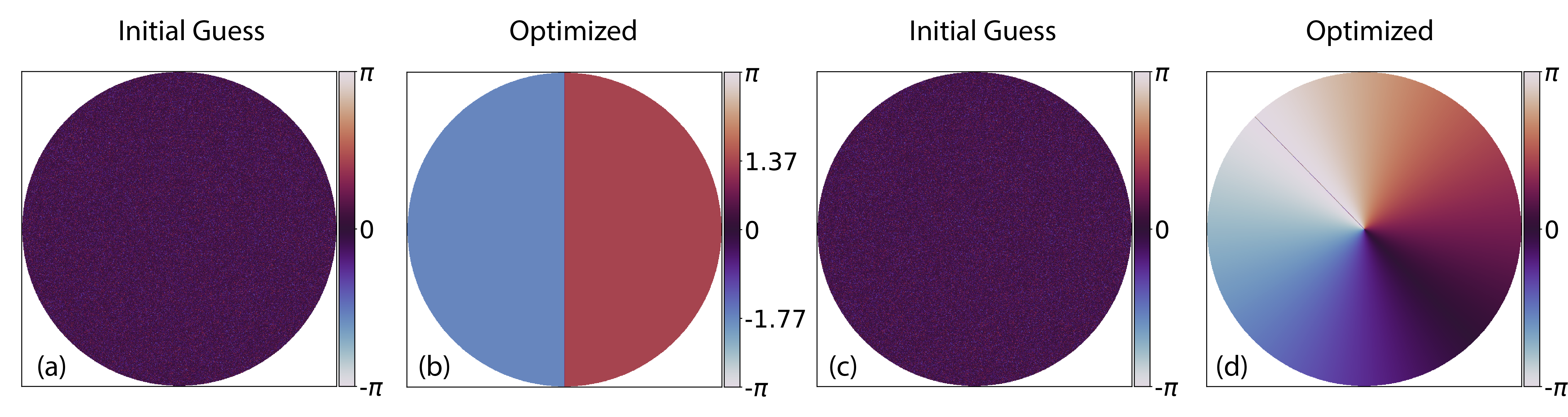}
    \caption{Numerical results of the projected power iterations. 
    (a)-(b): Initial guess (a) and final outcome (b) of the pupil of optimization for a single given direction $\n=(1, 0)$.
    (c)-(d): Initial guess (c) and final outcome (d) of the pupil of optimization averaged over all directions on the unit circle. 
    Both initial guesses (a) and (c) are random phase masks with values between $(-\pi)$ and $\pi$.}
    \label{fig: projected power iterations result}
\end{figure}

To extend our approach to 2D, we first choose a direction defined by a unit vector $\n$. The directional derivative $D_\n g(\r) \in \mathbb{C}$ of $g$ along a given direction $\n$ at $\r=\0$ is defined as
\begin{equation}
    \label{eq: directional derivative}
    D_\n g(\0) = \left(\underset{t \to 0}{\text{lim}}\left.\frac{g(\r + t\n) - g(\r)}{ t}\right)\right|_{\r=\0}. 
\end{equation}
From \eqref{eq: pupil constraints} and the properties of Fourier transform, 
\eqref{eq: directional derivative} can be expressed as
\begin{eqnarray}
    D_\n g(\0) &=& \F^{-1} \left\{\j (\k \cdot \n)P(\k) \right\}(\0) \\\label{eq:directional-gradient-ft}
    &=& 
    \frac{\j}{2\pi}\iint_\Lambda P(\k) (\k \cdot \n) \d \k. \label{eq: supp directional derivative in Fourier}
\end{eqnarray}   
where $\cdot$ denotes the scalar product between two vectors.
Our aim is to find the pupil function $P$ that maximizes the directional metric
\begin{equation}
    \label{eq: directional metric}
    m_\n(P) = \left|\iint_\Lambda P(\k) (\k \cdot \n) \d \k\right|^2.
\end{equation}
Using the triangular inequality and Eq. (\ref{eq: pupil constraints}), we have that
\begin{align}
    \nonumber
    m_\n(P) &\leq \left(\iint_\Lambda \left| P(\k) (\k \cdot \n) \right|\d\k \right)^2 \\\nonumber
    &= \left(\iint_\Lambda \left| P(\k)\right| \left| \k \cdot \n \right|\d\k \right)^2 \\
    &= \left(\iint_\Lambda  \left| \k \cdot \n \right|\d\k \right)^2.\label{eq: half-moon optimal demonstration}
\end{align} 
Equality is achieved when $P(\k) (\k \cdot \n) = \e^{\j \alpha}$ for a constant $\alpha$ for all $\k \in \Lambda$. 
This yields that $P(\k) = \e^{-\j\operatorname{Arg}(\k \cdot \n) + \alpha}$ and corresponds to a half-moon pupil function as depicted in Fig. \ref{fig: projected power iterations result}(b).
This half-moon pupil function is characterized by a $\pi$ phase jump between the two half disks.
It yields the optimal PSF that maximizes the gradient of the electric field in a single direction. 

If one wants to minimize a single-directional derivative norm, the optimal pupil function corresponds to a half-moon PSF. This supports the result of our CRB optimization as we obtain 2 PSFs that maximize gradients along orthogonal directions to solve our 2D localization-estimation problem. 

\subsection{Optimization Averaged Over All Directions}
Another desirable metric to optimize is the isotropic version of $m_\n$, averaging over all directions $\n$ on the unit circle $O$. This metric is
\begin{equation}
    \label{eq: isotropic metric}
    m_O(P) = \int_O \left|\iint_\Lambda P(\k) (\k \cdot \n) \d \k\right|^2 \d \n.
\end{equation}
This expression is less straightforward to maximize.
Still, we can reformulate it as a quadratic-programming problem in the Hilbert space $\mathcal{H}$ of compactly-supported functions $f: \Lambda \rightarrow \mathbb{C}$ with the Hermitian product $f^\mathrm{H}g = \iint_\Lambda f^*(\k) g(\k)\d\k$. For convenience, we will denote by $p \in \mathcal{H}$ the pupil function and, for any unit vector $\n$, $k_\n \in \mathcal{H}$ as
\begin{equation}
    \label{eq: def k scalar n hilbert}
    k_\n: \k \rightarrow (\k \cdot \n)^*.
\end{equation}
The metric to optimize can be rewritten as
\begin{align}
    \nonumber
    m_O(P) &= \int_O |k_\n^\mathrm{H} p|^2 \d\n \\\nonumber
    &= \int_O p^\mathrm{H} k_\n k_\n^\mathrm{H} p \d\n \\     
    &= p^\mathrm{H} \left(\int_O k_\n k_\n^\mathrm{H} \d\n\right) p \label{eq: exchange operator place} \\
    &= p^\mathrm{H} K p,
\end{align}
where we have introduced the symmetric linear operator $K = \int_O k_\n k_\n^\mathrm{H} \d\n$, operating in the Hilbert space $\mathcal{H}$.
To obtain Eq. (\ref{eq: exchange operator place}), the reordering of the integrals is legitimated by the compact support of functions in $\mathcal{H}$. 

Hence, we want to maximize this quadratic form over the set of pupil functions $p \in \mathcal{H}$ with the unit element-wise norm constraint of Eq. (\ref{eq: pupil constraints}). We compute the solution numerically using projected power iterations as
\begin{equation}
    \label{eq: projected power iterations}
    \begin{cases}
        p_0 : \k \rightarrow 1, \\
        p_{i+1} = \text{Proj}( K p_i ), \text{ for } i = 0, 1, \ldots.
    \end{cases}
\end{equation}
with the projector on the constraint set defined as:
\begin{equation}
    \label{eq: projector definition}
    \text{Proj}(f): \k \rightarrow \begin{cases}
        \frac{f(\k)}{|f(\k)|}, \quad & f(\k) \neq 0 \\
        1, \quad & \text{ otherwise.}
    \end{cases}
\end{equation}

We see in Fig. \ref{fig: projected power iterations result}(d) the result of these power iterations.
They yield the phase ramp phase characteristics of the donut PSF.
The integral over $\n \in O$ is discretized uniformly over the unit circle with $100$ terms.
The pupil function $p \in \mathcal{H}$ is discretized over a uniform grid for $(k_\text{x}, k_\text{y})$, as a $(2048\times2048)$ px image. 
A total of $100$ power iterations is performed. 

The projected power iterations employed for our complex-valued quadratic-programming problem with quadratic constraints lack theoretical convergence guarantees.
Similar formulations arise in other applications, such as electromagnetic beamforming \cite{he2022qcqp}.
However, by repeating the power iterations with different initial guesses, the donut pupil function consistently emerges as the solution, which demonstrates its robustness.

If one wants to minimize directional derivative norm averaged along all directions, the optimal pupil function corresponds to a donut PSF.
This supports the result of our CRB optimization as we obtain the donut PSF when only a single shifted PSF is allowed. 
\clearpage

\section{Supporting Figures and Tables}
\subsection{Additional Results of the First Group of Experiments in the Manuscript}
\begin{figure}[ht]
    \centering
    \includegraphics[width=0.8\textwidth]{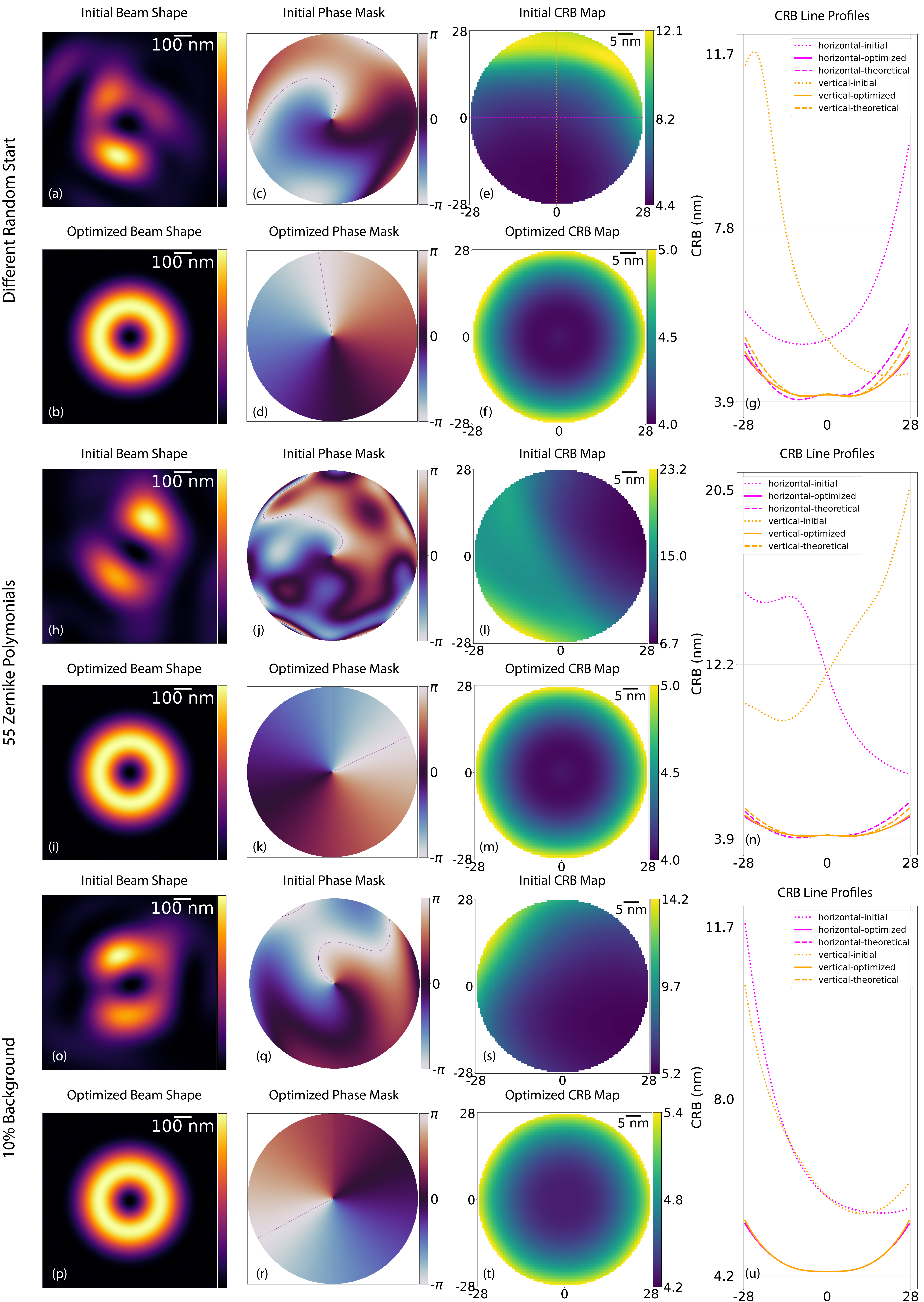}
    \caption{Optimization results of additional experiments with the setup of the first group of experiments with a single beam shape.
    Top two rows: Different random initialization of Zernike coefficients.
    Middle: First 55 (instead of 15) Zernike polynomials.
    Bottom two rows: 10\% background (instead of 5\%).
    (a)/(h)/(o) Initial beam shape.
    (b)/(i)/(p) Optimized shape.
    (c)/(j)/(q) and (d)/(k)/(r): Corresponding phase masks of the initial and optimized beams.
    (e)/(l)/(s) and (f)/(m)/(t): CRB map of the initial and the optimized. 
    (g)/(n)/(u) CRB profiles of the horizontal (represented in color magenta) and vertical (represented in color orange) lines through the center of the FOV for the initial (dotted), the optimized (solid), and the baseline (dashed).}
    \label{fig:perturbed-donut}
\end{figure}

In the first group of experiments, we consider the scenario of only one beam shape. 
In Fig. 2 of the manuscript, we presented the results with a donut perturbed by a phase mask composed of the first 15 Zernike polynomials with random coefficients with 5\% background as initial.
In Fig. \ref{fig:perturbed-donut}, we include the results of three additional experiments: another random start, another number of Zernike polynomials and another level of background. 
We see in Fig. \ref{fig:perturbed-donut} that, regardless of the various experimental conditions, the optimized beams converge consistently to a perfect donut beam, as indicated by the beam shape, the corresponding phase mask, and the good alignment of the CRB line profiles with the an ideal donut beam.
The evolution of the loss during the optimization process of all the experiments in this group in Fig. \ref{fig:loss} shows clear convergence, which confirms the optimality of the donut beam.
\clearpage
\subsection{Additional Results of the Second Groups of Experiments in the Manuscript}
\begin{figure}[ht]
    \centering
    \includegraphics[width=\textwidth]{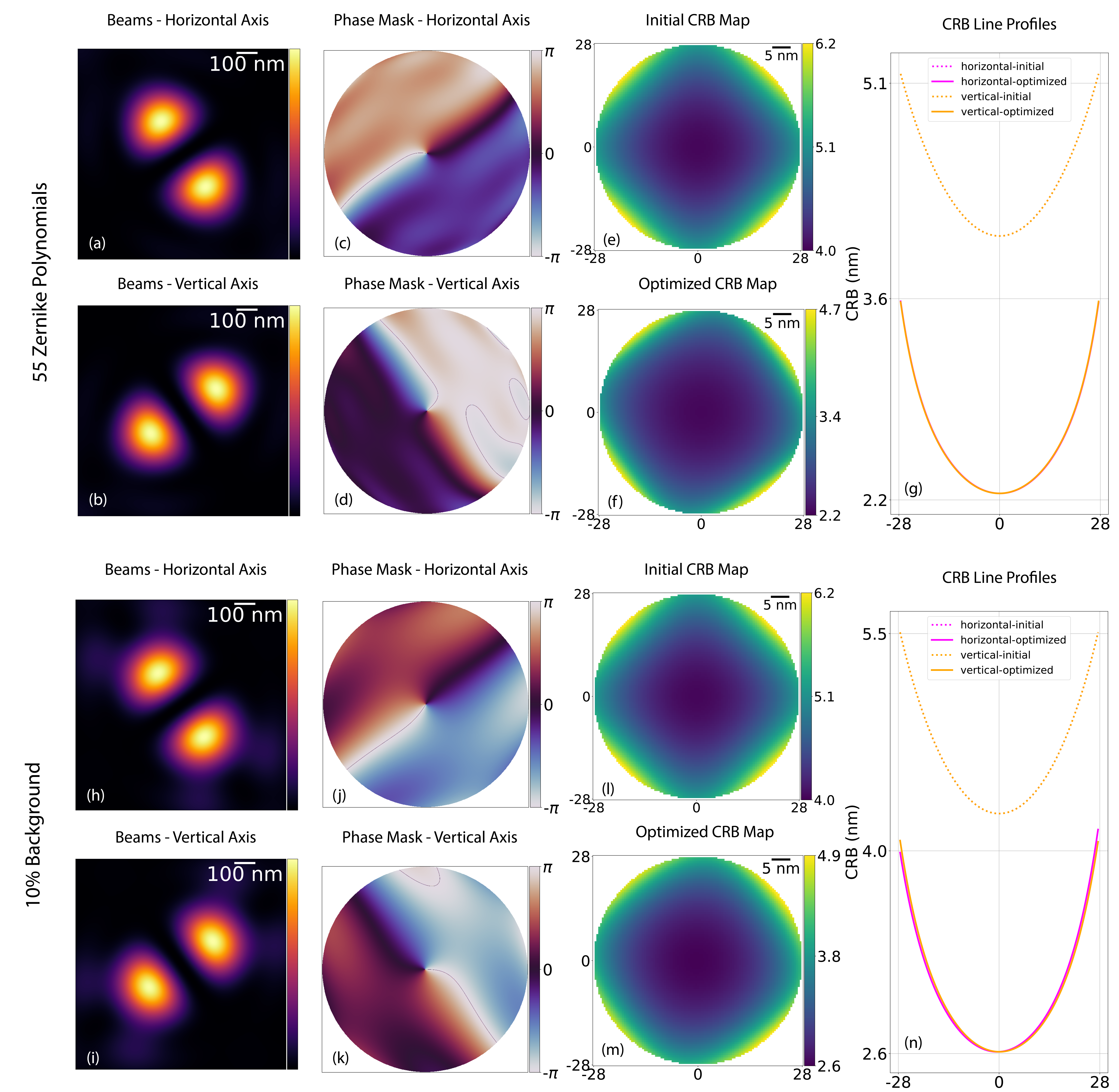}
    \caption{Optimization results of additional experiments with the setup of the second group of experiments with two pairs of beam shapes.
    First two rows: First 55 (instead of 15) Zernike polynomials.
    Last two rows: 10\% background (instead of 5\%).
    (a)/(h) The intensity of all the four optimized beams superimposed into one image.
    (b)/(i) CRB map of the optimized result.
    (c)/(j) and (d)/(k): Optimized beams at locations on the horizontal (b) and vertical (c) axis.
    (e)/(l) and (f)/(m): Corresponding phase masks of the two optimized beams.
    (g)/(n) CRB profiles of the horizontal (represented in color magenta) and vertical (represented in color orange) lines through the center of the FOV for the initial (dotted) and the optimized (solid).}
    \label{fig:halfmoon-additional}
\end{figure}

\begin{figure}[ht]
    \centering
    \includegraphics[width=\textwidth]{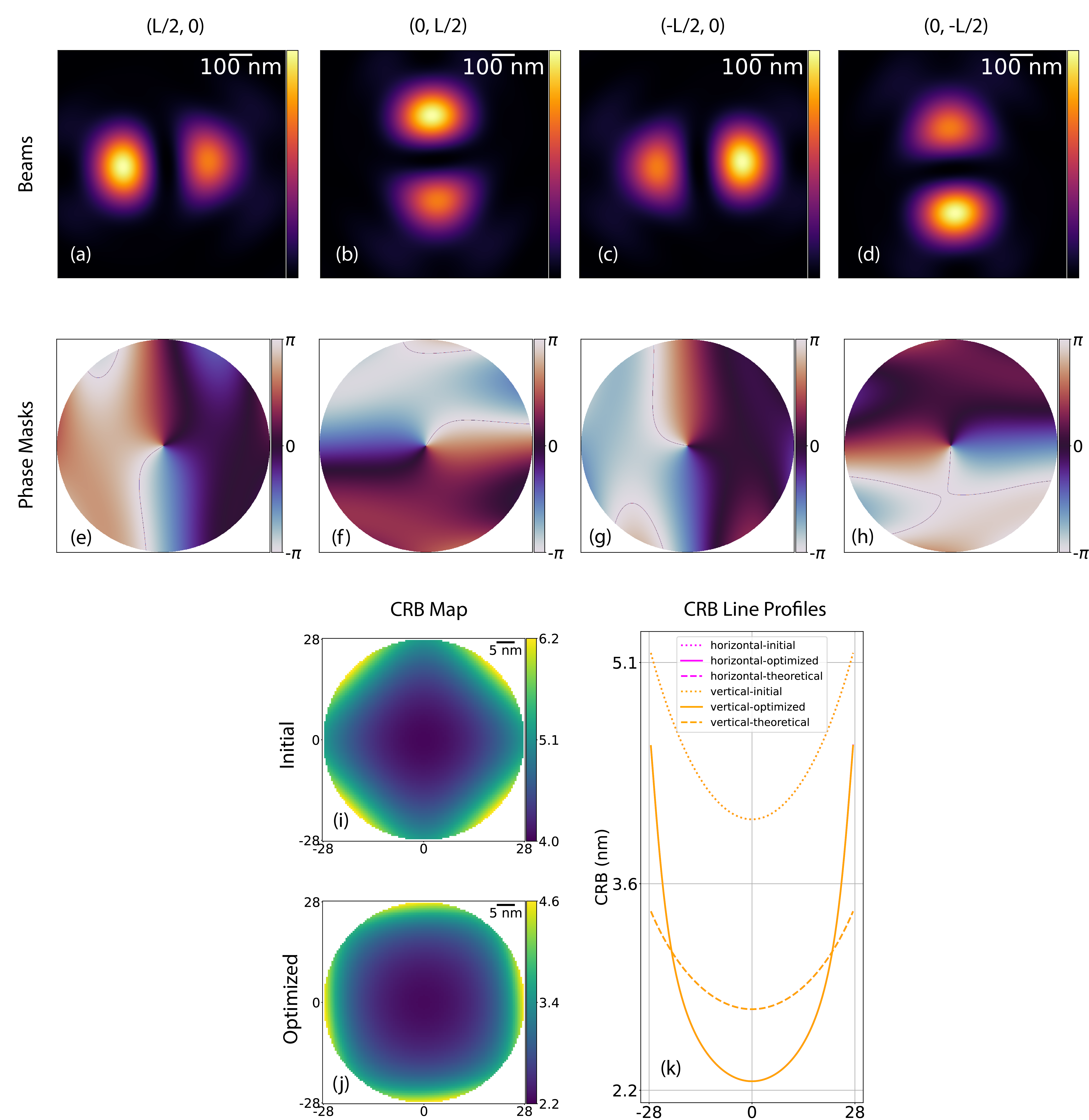}
    \caption{Optimization results with the setup of the second group of experiments with four independent beams(instead of in two groups thereof). 
    First row: Optimized beams centered at the four locations $(\frac{L}{2}, 0)$ (a), $(0, \frac{L}{2})$ (b), $(-\frac{L}{2}, 0)$ (c) and $(0, -\frac{L}{2})$ (d). 
    Second row: Associated phase masks of beams in the first row.
    (i) and (j): Initial and optimized CRB maps.
    (k) CRB line profiles along the horizontal and vertical straight lines (magenta and orange) through the center of the CRB maps for the initial (dotted), optimized (solid), and theoretical (dashed) cases.}
    \label{fig:4beams}
\end{figure}

In the second group of experiments, we consider the scenario of two pairs of beams arranged on a square. 
In Fig. 3 of the manuscript, we presented the results using the first 15 Zernike polynomials and 5\% background as initial start.
In Fig. \ref{fig:halfmoon-additional} and Fig. \ref{fig:4beams}, we include the results of three additional experiments: another number of Zernike polynomials, another level of background, and the case in which all the four beams are independent. 
We see in Fig. \ref{fig:halfmoon-additional} that, regardless of the various experimental conditions, the optimized beams are very robust and converge consistently to the half-moon-like shape.
The evolution of the loss during the optimization process of all the experiments in this group in Fig. \ref{fig:loss} shows clear convergence, which confirms the optimality of the half-moon beam.

\clearpage
\clearpage
\subsection{Other Supporting Tables and Figures}
In Table \ref{tab:pertured-donut-coeffs}, we compare the Zernike coefficients of the initial random start and the optimized result for the experiment presented in Fig. 2 of the manuscript.
Here, the additional term $\kappa$ on the phase mask - the vortex is treated as a vector in the basis alongside the first 15 Zernike polynomials. 
We see that the magnitude of the coefficient of the vortex is 100 times that of the others (except the first Zernike mode which is a constant 1), which confirms that the optimized beam is in fact almost a perfect donut beam.

\begin{figure}[ht]
    \centering
    \includegraphics[width=0.9\textwidth]{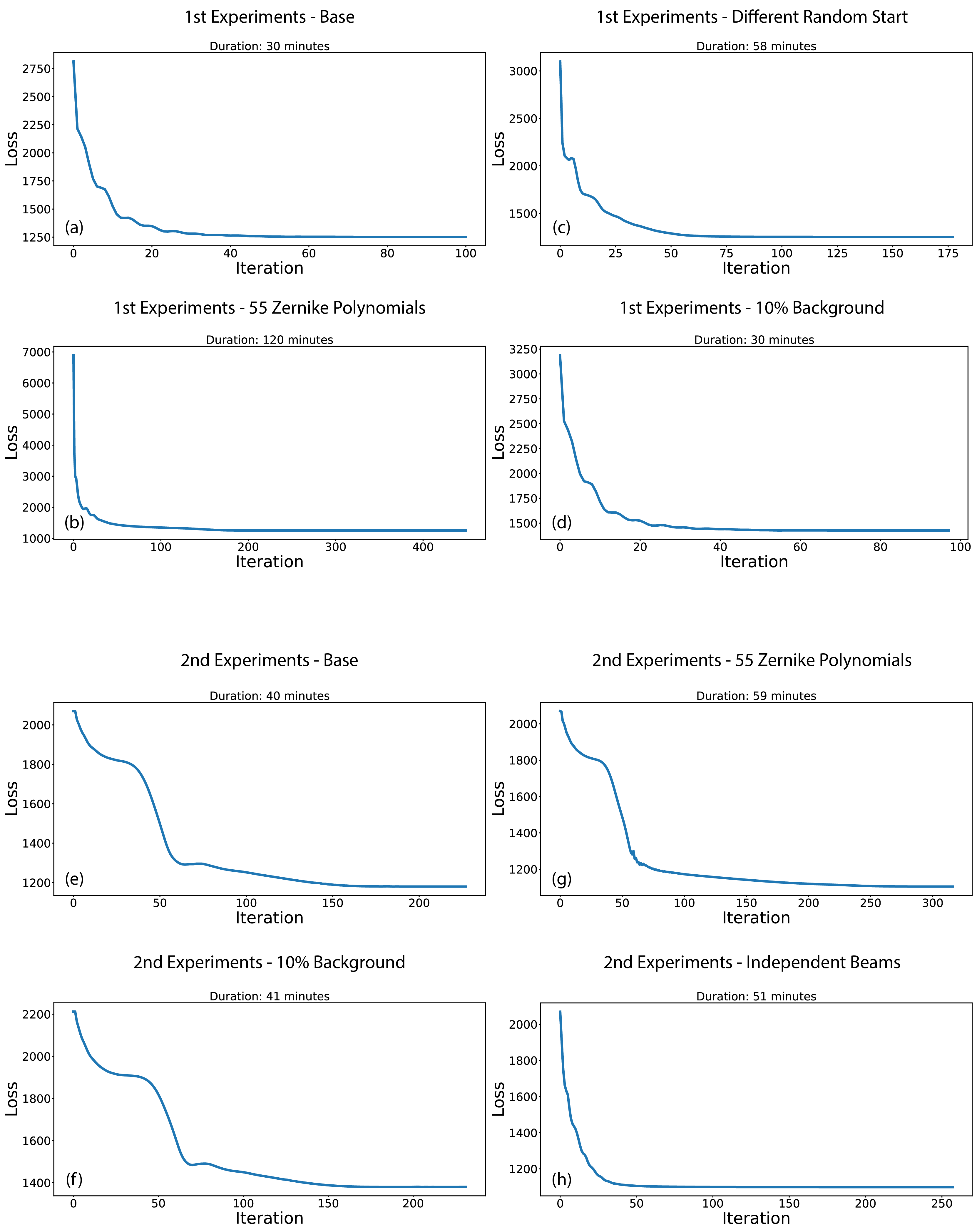}
    \caption{Loss and duration of the optimization algorithm.
    (a)-(d): First group of experiments of seven identical perturbed donut beams. 
    (a) Experiment presented in Fig. 2 in the manuscript.
    (b)-(d) Three additional experiments in Fig. \ref{fig:perturbed-donut}.
    (e) Experiment presented in Fig. 3 in the manuscript.
    (f)-(h) Experiments in Fig. \ref{fig:halfmoon-additional} ((f)-(g)) and Fig. \ref{fig:4beams} (h). }
    \label{fig:loss}
\end{figure}

In Fig. \ref{fig:loss}, we show the evolution of the loss and the run time of the optimization algorithm for all the experiments presented in the manuscript and the Supplement.
\begin{table}[ht]
    \centering
    \begin{tabular}{c|c|c}
    \hline
    Zernike Polynomial & Initial Coefficient & Optimized Coefficient \\
    \hline
      0& 0.19&   4.0264\\
      1&-0.11&  -0.0001\\
      2&-0.15&   0.0281\\
      3&-0.60&  -0.0129\\
       4&0.52&   0.0034\\
      5&-0.31&   0.0008\\
       6&0.28&  -0.0013\\
       7&0.27&  -0.0021\\
      8&-0.19&  -0.0073\\
      9&-0.52&  -0.0088\\
       10&0.65&   0.0094\\
       11&0.37&   0.0050\\
      12&-0.66&   0.0001\\
      13&-0.62&   0.0014\\
      14&-0.05&  -0.0024\\
      \hline
       Vortex &1.00&   0.9906\\
       \hline
    \end{tabular}
    \caption{Initial and optimized Zernike coefficients of the first 15 Zernike polynomials of the same beam shape shared by all seven beams in the first group of experiments. }
    \label{tab:pertured-donut-coeffs}
\end{table}

\twocolumn
\bibliographyfullrefs{bib}

\end{document}